\documentclass[twocolumn,aps,pra,superscriptaddress,showpacs,amsmath,amssymb,floatfix]{revtex4-1}

\usepackage{graphicx}
\usepackage{amssymb}
\usepackage{amsmath}
\usepackage{epsfig}
\usepackage{color}
\usepackage{mathtools}
\usepackage[colorlinks,linkcolor=blue,anchorcolor=blue,citecolor=blue,urlcolor=blue]{hyperref}
\usepackage[normalem]{ulem}

\begin{document}

\title{Quantum memory and non-demolition measurement of single phonon state with nitrogen-vacancy centers ensemble}
\author{Rui-xia Wang}
\author{Kang Cai}
\affiliation{Department of Physics, Tsinghua University, Beijing 100084, China}

\author{Zhang-qi Yin}\email{yinzhangqi@tsinghua.edu.cn}
\affiliation{Center for Quantum Information, Institute for Interdisciplinary Information Sciences, Tsinghua University, Beijing 100084, China}

 \author{Gui-lu Long}\email{gllong@tsinghua.edu.cn}
 \affiliation{Department of Physics, Tsinghua University, Beijing 100084, China}

\date{\today }

\begin{abstract}
In diamond, the mechanical vibration induced strain can lead to interaction between the mechanical mode and the nitrogen-vecancy (NV) centers. In this work, we propose to utilize the strain induced coupling for the quantum non-demolition (QND) single phonon measurement and memory in diamond. The single phonon in a diamond mechanical resonator can be perfectly absorbed and emitted by the NV centers ensemble (NVE) with adiabatically tuning the microwave driving.  An optical laser drives the NVE to the excited states, which have much larger coupling strength to the mechanical mode. By adiabatically eliminating the excited states under large detuning limit, the effective coupling between the mechanical mode and the NVE can be used for QND measurement of the single phonon state. Under realistic experimental conditions, we numerically simulate the scheme. It is found that the fidelity of the absorbing and emitting process can reach a much high value. The overlap between the input and the output phonon shapes can reach $98.57\%$.
\end{abstract}
 \maketitle

\section{Introduction}
In quantum information processing, one of the key challenges is to realize the
effective coupling, or communication, between distant qubits. Usually, the photon
is used as flying qubit for quantum communications \cite{DLCZ}, or inducing the effective coupling
between the distant qubits \cite{Yin2007}. For example, in superconducting quantum circuit systems,
the microwave photon confined in transmission lines resonator is used as a quantum
bus \cite{Song2017}. Recently, with the advances in the fabrication and manipulation of the mechanical systems \cite{QizheIn1,WangIn4,VladimirIn6,ConnellIn7}, an alternative way of using phonon to couple distant qubits was proposed \cite{SchuetzIn2}. The surface acoustic wave (SAW) has been successfully coupled with superconducting
qubits \cite{Patricio2016, Martin2014}. The advantages of the phonon quantum bus compared with the photon quantum bus are the high quality factor and small effective size \cite{Patricio2016, SchuetzIn2, painter2015}. The speed of SAW is $5$ orders of magnitude slower than the speed of light \cite{Martin2014, ManentiIn3}.
Therefore, the wavelength of SAW at GHz is also $5$ orders of magnitude smaller than the
microwave light. In this way, the individual superconducting qubit addressing by SAW is
possible. There are many schemes concerning single phonons, for example, the phonon states preparation \cite{Galland3,Matsuda5,Matsuda6}, detection \cite{Galland3,Matsuda5,Matsuda6,Yanay4,Chai7,Bai8,Woolley11,Ohm9},
and the phonon mediated interface  \cite{Neto10}.

As we know, in quantum optics experiments, the single photon detectors are widely used. For microwave photons, the quantum non-demolish (QND) measurement for
photon number states has been realized in both cavity QED and superconducting circuit QED systems \cite{Haroche2007}. The QND measurement in circuit QED systems
can be used for quantum error corrections. In order to further develop the phonon based quantum information processing,
the efficient single phonon detector is needed.  The ultimate goal is to realize QND measurement on single phonon state. Inspired by the QND measurement for photon number states, we use the strain induced phonon and the NVE coupling in the diamond to get the strong nonlinearity for the QND measurement of the single phonon state. The geometry of the mechanical crystal structure contains a diamond crystal with rectangular holes arranged periodically which can precisely manipulate mechanical vibrations \cite{painter2009}.

The nitrogen-vacancy (NV) center, which consists a substitutional nitrogen
atom and adjacent vacancy in diamond, is one of the most promising system
for solid-state quantum information processing. NV centers can be
controlled by microwave with long coherence time even at room temperature. The diamond
lattice vibration, or phonon, can couple
with the NV centers ensemble (NVE)  \cite{Ovartchaiyapong2014,Golter2016} and be cooled by the NVE \cite{MacQuarrie2017}. 
There are two different kind of mechanism for coupling the NV center and the phonons. The first on is magnetic field induced coupling \cite{Rabl2008, Rabl2010, du2009, zhou2009, yin2013, ma2016, cai2017}, and the second one is the strain induced coupling \cite{Bennett2013, wang2016x, wang2016l}. Here we focus on the second method. 
The strain induced coupling between NV centers electron spins in the ground state and the phonons is usually quite small, e.g. less than $1$ kHz  \cite{Bennett2013}. Recently, the excited-state electron-phonon coupling has been reported in experiments \cite{wang2016l}. It is about $6$ orders of magnitude stronger than the ground-state electron-phonon coupling \cite{maze2011,doherty2011}. By taking the advantage of strong coupling, the quantum control of the internal states of a NV center has been realized by using optomechanical sideband transition\cite{wang2016l}. We can use this strong coupling mechanism to induce the strong effective nonlinear coupling between the phonon and the NVE, and realized QND phonon number
measurement. 

In this work, the strain-induced spin-phonon coupling is employed in designing the scheme for the single phonon absorption, emission and QND measurement. We consider the NVE situated a few $\mu m$ below the diamond surface which is shown in Fig. \ref{scheme}(a). The coupling strength between the NVE and the single phonon can be enhanced by a factor $\sqrt{N}$ through the collective excitation, which can reach the strong coupling regime. The resonant frequency of the phonons propagating in the diamond can be controlled by the rectangular holes shown in the upper diagram of Fig. \ref{scheme}(a), which are periodically arranged on the diamond chip. The single phonon with the resonant frequency can be absorbed by the NVE, and the absorption will induce the resonant frequency shift which can be revealed in the phonon absorption spectrum, then we can detect the single phonon state through probing the resonance frequency shift of the phonons in the diamond.

The inverse process of the single phonon detection is a emission process.
The emitted phonon shape can be controlled by regulating the driving pulse
acting on the NVE. The similarity between the emitted and the absorbed
single phonon reaches $98.57\%$ if we inverse the driving pulse when emitting
compared with the absorbing process. If the single phonon state is in quantum state
before absorbing, the size and shape of the phonon will not be changed when
emitting, this process is a quantum non-demolition (QND) measurement process, which
realizes an ideal projective measurement that leaves the system in an
eigenstate of the phonon number \cite{Haroche2007,Braginsky14r1,Grangier14r2,Nogues14r3}.

\section{Model}

\begin{figure}[ht]
\begin{center}
\includegraphics[width=8cm,angle=0]{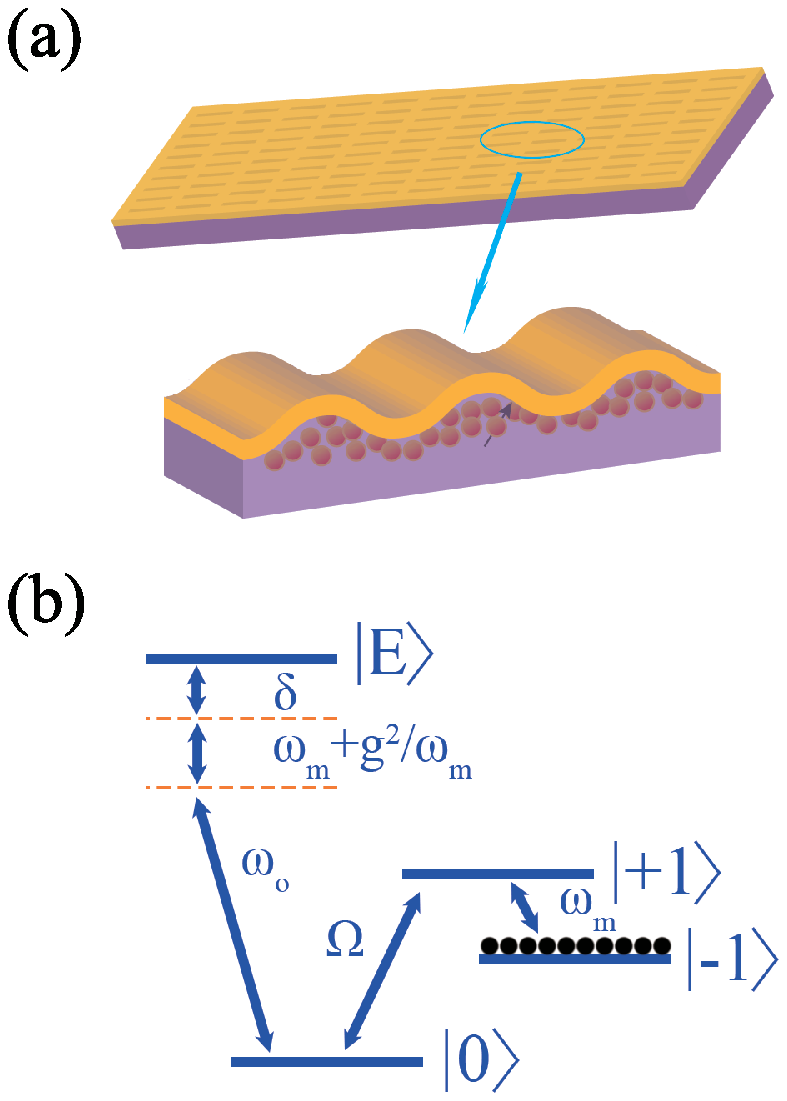}
\caption{(a) A schematic diagram of the phononic crystal. The NVE located near the surface coupling to a single phonon and a laser field. (b) The electronic structure of the NV center. $\Omega$ is the Rabi frequency between the energy levers $|0\rangle $ and $|+1\rangle $ induced by the microwave drive. $\omega_m$ is the phonon mode. $\omega_o$ is the optical driving frequency between the energy levers $|0\rangle $ and $|E\rangle $ inducing the Rabi frequency of $\Omega_o$.}\label{scheme}
\end{center}
\end{figure}

As schematically shown in Fig. \ref{scheme}(a), we consider a diamond chip with periodically arranged rectangle holes to adjust the refractive index for the surface acoustic wave (SAW). The NVE are located near the diamond surface coupling to the laser, microwave and SAW.

With a zero magnetic field, the spin-triplet ground state of the\ NV center
splits into two energy levels, $M_{s}=0$ and the nearly degenerate sublevels
$M_{s}=\pm 1$ \cite{Manson1r49}. We apply an external magnetic field $\overrightarrow{B}%
_{ext}$ along the crystalline direction $[100]$ of the NV center \cite{Saito12} to
split the degenerate sublevels $M_{s}=\pm 1$, which results in a three-level
system denoted by $|0\rangle =|^{3}A,M_{s}=0\rangle $, $|-1\rangle
=|^{3}A,M_{s}=-1\rangle $, $|+1\rangle =|^{3}A,M_{s}=+1\rangle $,
respectively. $|E\rangle$ is an excited state with an energy gap of about 1.189eV to the ground state.\cite{Rogers2008} The schematic diagram of the NV center energy structure is
shown in Fig. \ref{scheme}(b).

The NV centers exist in a diamond crystal and they can sense the lattice vibrations.
If there is a phonon produced in the lattice, the NV center can absorb it when the
mechanical frequency and the energy splitting of the NV centers are matched \cite{Bennett2013,Yin2015a}.
For this system, the phonon mode $a_{m}$ with the frequency $\omega _{m}$
couples to the NV centers with transition of $|-1\rangle $ to $|+1\rangle $. A classical
microwave field $\varepsilon \left( t\right) =\widetilde{\varepsilon }\left(
t\right) e^{-i\omega _{01}t}$\ drives the transition $|+1\rangle $ to $%
|0\rangle $ with a Rabi oscillation frequency $\Omega(t) $ ($%
\omega _{01}$ is the frequency splitting between the levels $|0\rangle \ $and%
$\ |+1\rangle $). The Rabi frequency $\Omega(t) $\ can be
written as $\Omega(t) =\Omega _{0}\alpha \left( t\right) $.
Within a good approximation, we assume that $\widetilde{\varepsilon }\left(
t\right) $ increase gradually from zero with $\widetilde{\varepsilon }\left(
0\right) \simeq 0$ to a finite strength. Because $\alpha(t)$ is proportional to $\widetilde{\varepsilon }(t)$ ($\alpha(t)
\propto \widetilde{\varepsilon}(t)$), then we can get $\Omega
\left( t\right) =\Omega _{0}^{\prime }\widetilde{\varepsilon }\left(
t\right) $, which discribes that the shape of the amplitude of the classical
driving microwave field totally decides the Rabi frequency.

Neglecting the dissipation of the NVE system, the Hamiltonian of the NVE
system including the coupling between the mode $a_{m}$ and the output can be
written as (setting $\hbar=1$) \cite{Duan2003,Yin2015}
\begin{eqnarray}
H =&&\sum^{N}_{i=1}G_m(|+1\rangle_i\langle-1|a_m+|-1\rangle_i\langle+1|a_m^{\dag})\nonumber
\\&&+\frac{\Omega(t)}{2}\sum^{N}_{i=1}(|0\rangle_i\langle+1|+|+1\rangle_i\langle0|)\nonumber
\\&&+i\sqrt{\kappa/2\pi}\int^{+\Delta\omega_e}_{+\Delta\omega_e}d\omega[a^{\dag}_me(\omega)-a_me^{\dag}(\omega)]\nonumber
\\&&+\int^{+\Delta\omega_e}_{+\Delta\omega_e}d\omega[\omega e^{\dag}(\omega)e(\omega)].
\end{eqnarray}

Where $G_{m}$\ is the coupling rate between single NV center and the phonon,
which is typically small. $e\left( \omega \right) $ denotes the
one-dimentional phonon modes which are free in the diamond and couple to the
phonon mode $a_{m}$. We need to consider the free propagating modes within a
finite bandwidth $\left[ \omega _{e}-\Delta \omega _{e},\omega _{e}+\Delta
\omega _{e}\right] $ that couple to the NV mode with the carrier frequency $%
\omega _{e}$. Within this bandwidth, the coupling between $e(\omega
)$ and the NV is a constant approximately which is denoted by $\sqrt{%
\kappa /2\pi }$\ for convenience. $\kappa $\ is the effective decay rate.

To obtain the relationship between the driving pulse shape and the output
phonon pulse shape, a simple picture by neglecting the
dissipation of the NVE system and the coupling of the mode $a_{m}$ to the
output is studied. Initially, the NVE are cooled to the ground state $|0\rangle $, and
a $\pi $ pulse is driven between the state $|0\rangle $ and $|-1\rangle $, and
then the NVE is in $|-1\rangle ^{\otimes N}$. We assume that the phonon
mode frequency $\omega _{m}$ equals to $\omega _{e}$ ($\omega _{e}$ is the
frequency splitting between the levels $|-1\rangle $ and $|+1\rangle $). In
the rotating frame, the Hamiltonian of the NVE system is
\begin{eqnarray}
H^{\prime}=&&\sum^{N}_{i=1}G_{m}(|+1\rangle_{i}\langle -1|a_{m}+|-1\rangle _{i}\langle +1|a_{m}^{\dag})\nonumber
\\&&+\frac{\Omega(t)}{2}\sum^{N}_{i=1}(|0\rangle\langle +1|+|+1\rangle\langle 0|).
\end{eqnarray}

We map the operator $|+1\rangle _{i}\langle -1|$ to the bosonic operators. $\sqrt{N}a^{\dag }=\sum_{i=1}^{N}|+1\rangle
_{i}\langle -1|$, $a^{\dag }a|n\rangle _{+1}=n|n\rangle _{+1}$ means there
are $n$ NVs that are in the state $|+1\rangle $. $d^{\dag}=|0\rangle _{i}\langle +1|$ and $d=|+1\rangle _{i}\langle 0|$ are the creation and annihilation operators for the ith NV center, and then the Hamiltonian can be written as follow, where $g_{m}=%
\sqrt{N}G_{m}$.
\begin{eqnarray}
H^{\prime }=g_{m}(a^{\dag }a_{m}+aa_{m}^{\dag })+\frac{\Omega (t)}{2}%
(d^{\dag }+d).
\end{eqnarray}

In the bases of $%
|N\rangle _{-1}|0\rangle _{+1}|0\rangle _{0}|1\rangle _{m}$, $|N-1\rangle
_{-1}|1\rangle _{+1}|0\rangle _{0}|0\rangle _{m}$, $|N-1\rangle
_{-1}|0\rangle _{+1}|1\rangle _{0}|0\rangle _{m}$, where the state $%
|p\rangle _{-1}|q\rangle _{+1}|r\rangle _{0}|s\rangle _{m}$ represents that the
numbers of the\ NVs which are in the states $|-1\rangle $, $|+1\rangle $,
and $|0\rangle $ are $p$, $q$ and $r$ respectively, and the number of phonons
in the diamond is $s$. The Hamiltonian can be written as a matrix as follow
in the three bases,
\begin{eqnarray}
H^{\prime }=\left[
\begin{array}{ccc}
0 & g_{m} & 0 \\
g_{m} & 0 & \frac{\Omega(t)}{2} \\
0 & \frac{\Omega (t)}{2} & 0\\
\end{array}
\right].
\end{eqnarray}

This Hamiltonian has the well-known dark state $|D\rangle $ with the form $%
|D\rangle =\frac{-\Omega(t)/2}{\sqrt{g_{m}^{2}+\left[ \Omega(t)/2%
\right] ^{2}}}|N\rangle _{-1}|0\rangle _{+1}|0\rangle _{0}|1\rangle _{m}+%
\frac{g_{m}}{\sqrt{g_{m}^{2}+\left[ \Omega(t)/2\right] ^{2}}}%
|N-1\rangle _{-1}|0\rangle _{+1}|1\rangle _{0}|0\rangle _{m}$. The dark
state means, the state will remain on the state $|D\rangle $, if the driving
pulse $\Omega(t)$ and the coupling strength satisfy the relationship $%
\cos \theta =\frac{\Omega(t)/2}{\sqrt{g_{m}^{2}+\left[ \Omega(t)/2%
\right] ^{2}}}$ and $\sin \theta =\frac{g_{m}}{\sqrt{g_{m}^{2}+\left[ \Omega
(t)/2\right] ^{2}}}$.

The effects of non-zero dissipation will be analyzed in section \uppercase\expandafter{\romannumeral4} where
the phonon detecting efficiency is calculated as well as the overlap of the absorbing and
emitting a single phonon with the real lossy system.

\section{Pulse shape}

To obtain the relationship of the pulse shape between the driving and the
phonon, we first consider the emitting process from state $|N-1\rangle _{-1}|0\rangle _{+1}|1\rangle _{0}|0\rangle _{m}$ to $|N\rangle _{-1}|0\rangle _{+1}|0\rangle _{0}|1\rangle _{m}$ under ideal
conditions. The Hamiltonian of the system can be written as
\begin{eqnarray}
H&=&g_{m}\left( a^{\dagger }a_{m}+aa_{m}^{\dagger }\right) +\frac{\Omega
\left( t\right) }{2}(d^{\dag }+d)\nonumber
\\
&&+i\sqrt{\kappa /2\pi }\int_{-\Delta \omega _{e}}^{+\Delta \omega
_{e}}d\omega \left[ a_{m}^{\dagger }e\left( \omega \right) -a_{m}e^{\dagger
}\left( \omega \right) \right]\nonumber
\\&&+\int_{-\Delta \omega _{e}}^{+\Delta \omega
_{e}}d\omega \left[ \omega e^{\dagger }\left( \omega \right) e\left( \omega
\right) \right].
\end{eqnarray}

Assume that, at the time $t=0$, the Raby frequency $\Omega \left( t\right)
=0 $, the NVE system are in the state $|N-1\rangle _{-1}|0\rangle _{+1}|1\rangle _{0}|0\rangle _{m}$, after
applying a classical driving pulse $\varepsilon \left( t\right) $, the Rabi
frequency, which is proportional to the intensity of the driving pulse,
changes slowly and is within the adiabatic approximation. Then we can
expend the state $|\Psi \rangle $ of the whole system into the
following form \cite{Duan2003}
\begin{eqnarray}
|\Psi\rangle=c_{d}|D\rangle\otimes|\phi _{0}\rangle +|N\rangle _{-1}|0\rangle _{+1}|0\rangle _{0}|0\rangle _{m}\otimes |\phi _{1}\rangle,
\end{eqnarray}
where $|\phi _{0}\rangle $ denotes the vacuum state of mode $e\left( \omega
\right) $ in the diamond, and
\begin{eqnarray}
|\phi _{1}\rangle =\int_{-\Delta \omega _{e}}^{+\Delta \omega _{e}}d\omega c_{\omega }e^{\dagger }\left( \omega \right) |\phi _{0}\rangle,
\end{eqnarray}
represents the single phonon output state. Initially, $c_{d}=1$, $c_{\omega
}=0$ and $\Omega \left( 0\right) =0$. Within the adiabatic approximation,
we would like to calculate the time evolution of the whole NVE system state $%
|\Psi \rangle $ by substituting it into the schr$\ddot{o}$dinger equation $i\partial
_{t}|\Psi \rangle =H|\Psi \rangle $. The coefficients $c_{d}$ and $c_{\omega
}$ can be got, which satisfy the following evolution equations:
\begin{eqnarray}
\dot{c_{d}}&=&\left( -\sqrt{\kappa /2\pi }\cos \theta \right)\int_{-\Delta \omega _{e}}^{+\Delta \omega _{e}}c_{\omega }d\omega,
\\
\dot{c_{\omega }}&=&-i\omega c_{\omega }+\sqrt{\kappa /2\pi }c_{d}\cos \theta.
\end{eqnarray}

We can get the solution of $c_{\omega }$ as follow
\begin{eqnarray}
c_{\omega }\left( t\right) =\sqrt{\kappa /2\pi }\int_{0}^{t}e^{-i\omega
\left( t-\tau \right) }c_{d}\left( \tau \right) \cos \theta \left( \tau
\right) d\tau,
\end{eqnarray}%
substituting the solution into the equation of $c_{d}$, leads to%
\begin{eqnarray}
\dot{c_{d}}=-\frac{\kappa \cos \theta }{\pi }\int_{0}^{t}\frac{%
\sin \left[ \delta \omega \left( t-\tau \right) \right] }{\left( t-\tau
\right) }c_{d}\left( \tau \right) \cos \theta \left( \tau \right) d\tau.
\end{eqnarray}

The bandwidth $\delta \omega $ satisfies $\delta \omega T\gg 1$, where the
operation time $T$\ characterizes the time scale for a significant change of
$c_{d}$ and $\sin \theta $, so the above function satisfies a $\delta $
function
\begin{eqnarray}
\delta \left( x\right) =\lim_{k\to\infty}\frac{1}{\pi }%
\frac{\sin kx}{x},
\end{eqnarray}%
then we can obtain%
\begin{eqnarray}
\dot{c_{d}}=-\frac{\kappa }{2}c_{d}\left( t\right) \cos
^{2}\theta,
\end{eqnarray}

The solution of $c_{d}$\ and $c_{\omega }$ are
\begin{eqnarray}
c_{d}=e^{-\frac{\kappa }{2}\int_{0}^{t}\cos ^{2}\theta \left( \tau \right)
d\tau },
\end{eqnarray}
\begin{eqnarray}
c_{\omega }\left( t\right) &=&\sqrt{\kappa /2\pi }\int_{0}^{t}e^{-i\omega
\left( t-\tau \right) }e^{-\frac{\kappa }{2}\int_{0}^{t}\cos ^{2}\theta
\left( \tau \right) d\tau }\nonumber
\\&&\cdot cos \theta \left( \tau \right) d\tau.
\end{eqnarray}

The single-phonon pulse shape $f\left( t\right) $ can be obtained by the
Fourier transformation
\begin{eqnarray}
f\left( t\right) &=&\frac{1}{\sqrt{2\pi }}\int_{-\Delta \omega
_{e}}^{+\Delta \omega _{e}}d\omega c_{\omega }\left( T\right) e^{-i\omega
\left( t-T\right) } \\
&=&\sqrt{\kappa }\cos \theta \left( t\right) e^{-\frac{\kappa }{2}
\int_{0}^{t}\cos ^{2}\theta \left( \tau \right) d\tau }.
\end{eqnarray}

If the Rabi frequency $\Omega \left( t\right) $, which is proportional to
the strength of the driving pulse,\ satisfies the function as $\Omega \left(
t\right) =2g_{m}\exp \left[ \kappa \left( t-\frac{T}{2}\right) /2\right] $,
the output phonon shape will be symmetric. The output
phonon shape function can be figured out as
\begin{eqnarray}
f\left( t\right) =\frac{\sqrt{\kappa }\sqrt{1+\exp
\left( -\frac{\kappa T}{2}\right) }}{\exp \left[ -\kappa \left( t-\frac{T}{2}%
\right) /2\right] +\exp \left[ \kappa \left( t-\frac{T}{2}\right) /2\right] },
\end{eqnarray}
the results is shown in Fig. 3 for $\kappa/2\pi =1.6\times 10^{5}$, $%
T=\frac{20}{\kappa }$. We assume that, in the absorption process, we know
the phonon shape in advance, such as the shape in Fig. 2. As the absorption
process is the time reversal of the emission process, and the phonon shape
is symmetrical in the time domain, if we reverse the temporal driving pulse
shape on NV centers in the absorbing process, the phonon can be completely
absorbed by the NVE \cite{Cirac13}. If the phonon shape is not symmetrical in the
time domain, it can also be completely absorbed as long as it is reversed
simultaneously with the driving pulse.

\begin{figure}[ht]
\begin{center}
\includegraphics[width=8cm,angle=0]{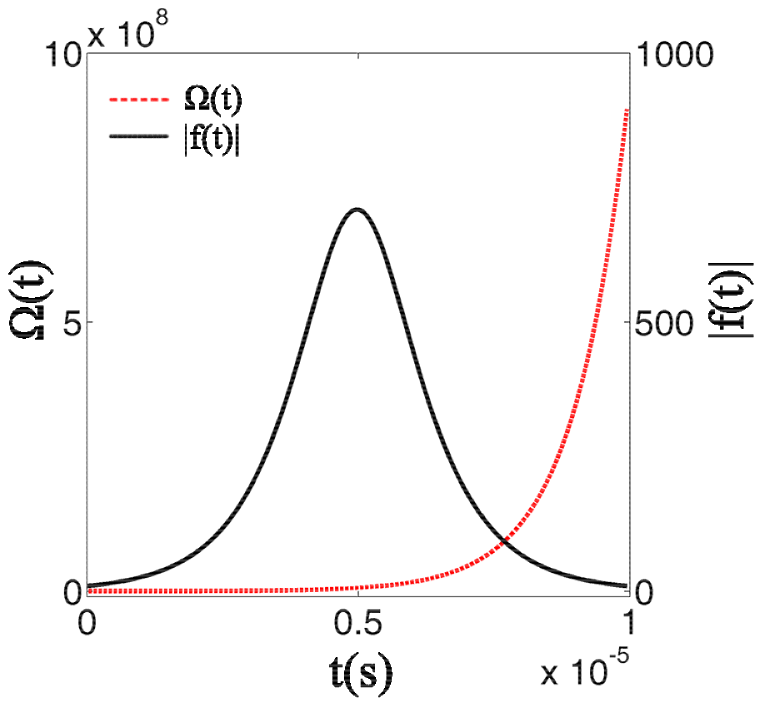}
\caption{The situation of the symmetric output phonon shape. The red line is the driving pulse shape applying to the NVE, and the black line represents the pulse shape of the output phonon.}\label{fig_robustness2}
\end{center}
\end{figure}

\section{QND measurement}

As schematically shown in Fig. 1, we consider the energy levels $|0\rangle$ and $|E\rangle$. After the absorption process, we stop the driving pulse $\Omega$ between the state $|0\rangle$ and $|+1\rangle$ and start the driving pulse $\Omega_o$ between the state $|0\rangle$ and $|E\rangle$ to detect the single phonon state.

The Hamiltonian of the detecting system is
\begin{eqnarray}
H&=&\omega _{m}a_{m}^{\dag }a_{m}+\omega _{0E}|E\rangle \langle E|\nonumber
\\
&&+\frac{\Omega _{o}}{2}\left( e^{i\omega _{o}t}|0\rangle \langle E|+e^{-i\omega_{o}t}|E\rangle \langle 0|\right)\nonumber
\\
&&+g\left( a_{m}^{\dag }+a_{m}\right)|E\rangle \langle E|
\end{eqnarray}
Applying the Schrieffer-Wolff transformation
\begin{equation*}
U=\exp [ \frac{g}{\omega _{m}}\left( a_{m}^{\dag }-a_{m}\right)
|E\rangle \langle E|]
\end{equation*}
to the Hamiltonian gives
\begin{eqnarray}
\widetilde{H}&=&UHU^{\dag}\nonumber
\\
&=&\omega _{m}a_{m}^{\dag }a_{m}-\frac{g^{2}}{\omega _{m}}|E\rangle \langle
E|+\omega _{0E}|E\rangle \langle E|\nonumber
\\
&&+\frac{\Omega _{o}}{2}[e^{i\omega _{o}t-\frac{g}{\omega _{m}}(a_{m}^{\dag }-a_{m})}|0\rangle\langle E|+H.c.]
\end{eqnarray}

Apply the RWA, and set $H_{0}=\omega _{m}a_{m}^{\dag }a_{m}+( \omega _{0E}-\frac{g^{2}}{\omega _{m}}-\delta) |E\rangle \langle E|$, the Hamiltonian becomes
\begin{eqnarray}
\widetilde{H}_r &=&\delta |E\rangle \langle E|+\frac{g\Omega
_{o}}{2\omega_m}a_{m}^{\dagger }|E\rangle \langle 0|+\frac{g\Omega _{o}}{2\omega_m}%
a_{m}|0\rangle \langle E|
\end{eqnarray}
Where $\omega _{0E}=\delta+\omega _{m}+\frac{g^{2}}{\omega _{m}}+\omega _{o}$.

Then we apply another RWA, and set $H_{0}^{\prime}=\delta |E\rangle \langle E|$, the Hamiltonian is calculated as
\begin{eqnarray}
\widetilde{H}_{r}^{^{\prime }}=\frac{g\Omega _{o}}{2\omega_m}e^{i\delta
t}a_{m}^{\dagger }|E\rangle \langle 0|+\frac{g\Omega _{o}}{2\omega_m}%
e^{-i\delta t}a_{m}|0\rangle \langle E|
\end{eqnarray}

When the interaction strength of $\frac{\Omega _{o}}{2}$ is far small compared with the detuning $\delta$, the time-average effective Hamiltonian is obtained as\cite{James2}
\begin{eqnarray}
\widetilde{H}_{eff}^{^{\prime }}&=&\frac{g^2\Omega _{o}^{2}}{4\omega_m^2\delta }%
a_{m}^{\dagger }a_{m}( |E\rangle \langle E|-|0\rangle \langle 0|)\nonumber
\\
&&-\frac{g^2\Omega _{o}^{2}}{4\omega_m^2\delta }|0\rangle \langle 0|
\end{eqnarray}

From the Hamiltonian we can see that the resonant frequency of the phonon crystal will be affected by the state of NVE, which means the resonant frequency shift is proportional
to the excitation number in state $|0\rangle $. When there is one NV center absorbing a single phonon, the frequency shift of the phonon crystal is
\begin{eqnarray}
\Delta f_{s}=\frac{g^2\Omega _{o}^{2}}{2\omega_m^2\delta }.
\end{eqnarray}

The effective dissipation of the excited state $|E\rangle$ is $\gamma_e=(\frac{\overline{g}}{\delta})^2\gamma_E$, where $\overline{g}=\frac{g\Omega _{o}}{2\omega_m}$ is the equivalent coupling strength, $\gamma_E$ is the decay rate of the energy level $|E\rangle$ with the population of 1. If we set $g=2\pi\times5MHz$, $\Omega_o=2\pi\times290MHz$, $\omega_m=2\pi\times900MHz$ \cite{wang2016l}, $\gamma_E=2\pi\times3MHz$ \cite{Robledo}, and $\delta=2\pi\times30MHz$, the effective dissipation and the frequency shift can be calculated as $\gamma_e=2\pi\times 2.16kHz$ and $\Delta f_{s}=2\pi\times 43.26kHz$ respectively, which implies that the single phonon absorption induced frequency shift can be detected distinctively. In fact, as long as we have $\delta\gg\gamma_E$, the demand for the resolution can be satisfied, and the scheme can be realized under the realistic experimental conditions.

\section{Dissipation effects}
In the above sections, the dynamical evolution has been analysed in ideal condition and the relationship between
the driving pulse shape and the emitting phonon shape has been obtained,
however, there are various losses in the real systems. If we consider the
dynamical evolution of the NVE system with the conditional Hamiltonian which
include the possible losses, the whole conditional Hamiltonian has the following form:
\begin{eqnarray}
H_{c} &=&-i\frac{\gamma _{1}}{2}|N-1\rangle _{-1}|1\rangle _{+1}|0\rangle _{0}|0\rangle _{m}\nonumber
\\&&-i\frac{\gamma _{0}}{2}|N-1\rangle _{-1}|0\rangle _{+1}|1\rangle _{0}|0\rangle _{m}-i\frac{\gamma _{m}}{2}a_{m}^{\dagger }a_{m}\nonumber
\\&&+g_{m}(a^{\dagger }a_{m}+aa_{m}^{\dagger })+\frac{\Omega (t)}{2}%
(d^{\dag }+d)\nonumber
\\&&+i\sqrt{\kappa /2\pi }\int_{-\Delta \omega _{e}}^{+\Delta \omega
_{e}}d\omega [ a_{m}^{\dagger }e(\omega) -a_{m}e^{\dagger
}( \omega)]\nonumber
\\&&+\int_{-\Delta \omega _{e}}^{+\Delta \omega
_{e}}d\omega[\omega e^{\dagger }(\omega)e( \omega
)].
\end{eqnarray}
Where $\gamma _{0}$ is the dephasing rate in state $|0\rangle $, $\gamma _{1}$ denotes the spontaneous emission of
state $|+1\rangle $, and $\gamma _{m}$ denotes
the phonon dissipation in the single NV evolution.

For numerical simulations, we need to discretize the field $e( \omega
) $ by introducing a finite but small frequency interval $\delta
\omega $ between two adjacent mode frequency. Then the number of the total
modes we have is $n=\frac{2\Delta \omega _{e}}{\delta \omega }+1$, the
frequency of the $jth$ mode $\omega _{j}$ which is denoted by $e_{j}$ is
given by\ $\omega _{j}=\left( j-\frac{n+1}{2}\right) \delta \omega $. The
Hamiltonian can be transformed to the form:
\begin{eqnarray}
H_{c} &=&-i\frac{\gamma _{1}}{2}|N-1\rangle _{-1}|1\rangle _{+1}|0\rangle _{0}|0\rangle _{m}\nonumber
\\&&-i\frac{\gamma _{0}}{2}|N-1\rangle _{-1}|0\rangle _{+1}|1\rangle _{0}|0\rangle _{m}-i\frac{\gamma _{m}}{2}a_{m}^{\dagger }a_{m}\nonumber
\\&&+g_{m}(a^{\dagger }a_{m}+aa_{m}^{\dagger })+\frac{\Omega (t)}{2}%
(d^{\dag }+d)\nonumber
\\&&+i\kappa _{e}\sum_{j=1}^{n}\sqrt{\delta \omega }[
a_{m}^{\dagger }e_{j}-a_{m}e_{j}^{\dagger }]\nonumber
\\&&+\sum_{j=1}^{n}(
\omega _{j}\delta \omega e_{j}^{\dagger }e_{j}),
\end{eqnarray}%
where $\kappa _{e}=\sqrt{\kappa \delta \omega /2\pi }$. Similarly, the state $|\Psi \rangle $ of the NVE system can be expanded by the discretized phonon pulse
state with the form $|\phi _{1}\rangle =\sum_{j=1}^{n}\delta \omega
b_{j}e_{j}^{\dagger }|vac\rangle $, which is
\begin{eqnarray}
|\Psi >&=&( c_{1}|N\rangle _{-1}|0\rangle _{+1}|0\rangle _{0}|1\rangle _{m}+c_{2}|N-1\rangle _{-1}|1\rangle _{+1}|0\rangle _{0}|0\rangle _{m}\nonumber
\\&&+c_{3}|N-1\rangle _{-1}|0\rangle _{+1}|1\rangle _{0}|0\rangle _{m})\otimes|\phi_{0}\rangle\nonumber
\\&&+|N\rangle _{-1}|0\rangle _{+1}|0\rangle _{0}|0\rangle _{m}\otimes|\phi _{1}\rangle.
\end{eqnarray}
Substituting $|\Psi \rangle $ into the schr$\ddot{o}$dinger equation $i\partial
_{t}|\Psi \rangle =H|\Psi \rangle $, we can get
\begin{eqnarray}
\dot{c}_{1}=-\frac{\gamma _{m}}{2}c_{1}-ig_{m}c_{2}+\kappa
_{e}\sum_{j=1}^{n}\widetilde{b}_{j},
\end{eqnarray}
\begin{eqnarray}
\dot{c}_{2}=-\frac{\gamma _{1}}{2}c_{2}-ig_{m}c_{1}-i\frac{%
\Omega \left( t\right) }{2}c_{3},
\end{eqnarray}
\begin{eqnarray}
\dot{c}_{3}=-\frac{\gamma _{0}}{2}c_{3}-i\frac{\Omega \left(
t\right) }{2}c_{2},
\end{eqnarray}
\begin{eqnarray}
\dot{b}_{j}=-\kappa _{e}c_{1}-i\widetilde{b}_{j}\omega _{j},
\end{eqnarray}%
where $\widetilde{b}_{j}=\sqrt{\delta \omega }b_{j}$. We assume that,
initially, there is a single phonon state in the diamond, which shape is
shown as Fig. 2, and all of the NV centers are in the state $|-1\rangle $,
which means $c_{1}=c_{2}=c_{3}=0$, $\sum \left\vert b_{j}\right\vert
^{2}\neq 0$. If we would like to absorb the phonon effectively, the driving pulse $%
\Omega ^{^{\prime }}\left( t\right) $ should be the time reversal of $\Omega
\left( t\right) $ in Fig. 2, that is $\Omega ^{^{\prime }}\left( t\right)
=g_{m}\exp \left[ \kappa \left( -t+\frac{T}{2}\right) /2\right] $, where $%
T=20/\kappa $. The shape control of the driving microwave pulse can be
easily achieved by modulating an arbitrary wave generator. We set $\gamma
_{0}/2\pi =0.16kHz$, $\gamma _{1}/2\pi =0.16kHz$ \cite{Gill2013}, $\gamma _{m}/2\pi =0.16kHz$ \cite{Tao2014,Ovartchaiyapong2014}, $g_{m}/2\pi =0.96MHz$, and $%
\kappa/2\pi =0.32MHz$. The numerical simulation results of absorbing process is
shown in Fig. 3a with the black lines. Under the condition of strong coupling between the NVE and
the phonon, the process is quasi-adiabatic, the value of $c_{2}$ is
distinctly smaller than $c_{1}$ and $c_{3}$. The first half is the process
that the free single phonon entering the strong coupling area with the NVE plays a
leading role, and the later half is predominantly the absorbing process. at
time $t=T$, the fidelity between the actual state and the ideal state is $%
99.38\%$. As the dissipations increase, the fidelity decreases obviously which is also shown in Fig. 3a with the blue lines. %
The red lines in Fig. 3a implies that, it will also reduce the fidelity of the state if the coupling strength and the effective %
decay rate $\kappa$ do not satisfy the adiabatic condition.

\begin{figure}[ht]
\begin{center}
\includegraphics[width=7.5cm]{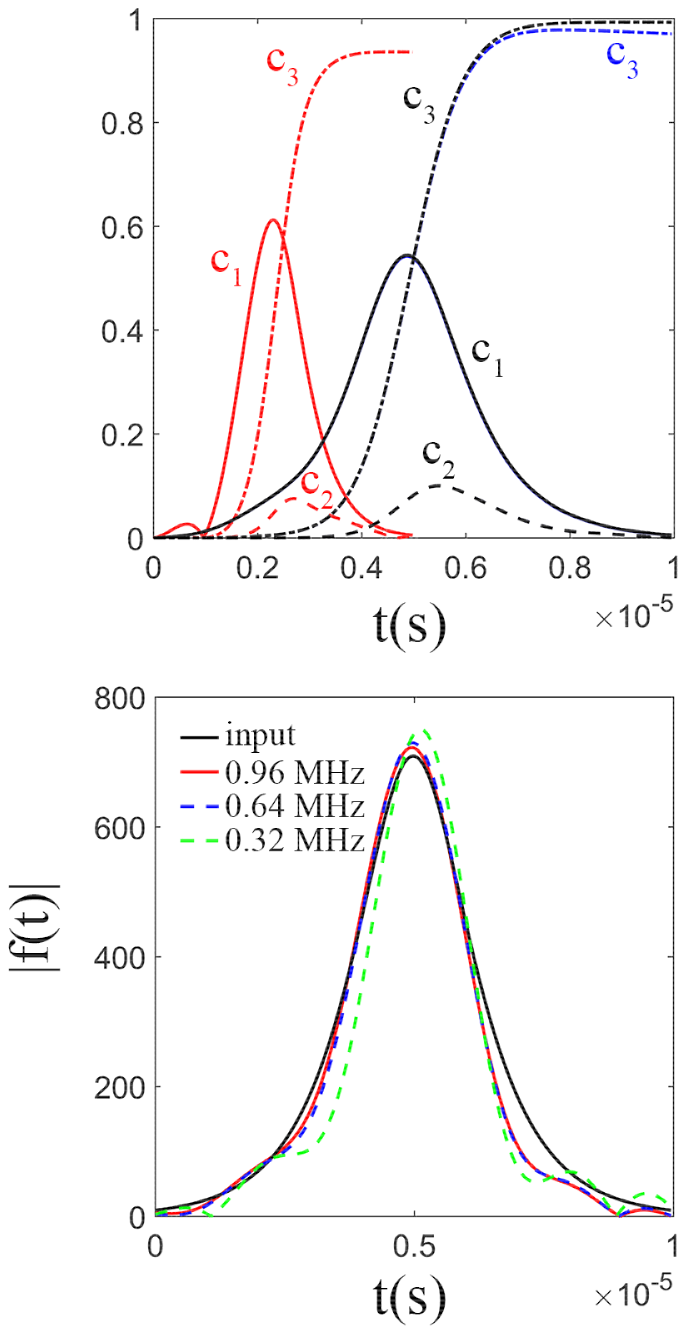}
\caption{The numerical simulation results. (a) is the absorbing process. The black lines represents the process with $\gamma_{0}/2\pi =0.16kHz$, $\gamma _{1}/2\pi =0.16kHz$, $\gamma _{m}/2\pi=0.16kHz$, $g_{m}/2\pi =0.96MHz$, and $\kappa/2\pi =0.32MHz$. The blue lines show the process with $\gamma_{0}/2\pi =1.6kHz$, $\gamma _{1}/2\pi =1.6kHz$, $\gamma _{m}/2\pi =1.6kHz$, $g_{m}/2\pi =0.96MHz$, and $\kappa/2\pi =0.32MHz$. The red ones denote the process with $\gamma_{0}/2\pi =0.16kHz$, $\gamma _{1}/2\pi =0.16kHz$, $\gamma _{m}/2\pi =0.16kHz$, $g_{m}/2\pi =1.92MHz$, and $\kappa/2\pi =0.64MHz$.
(b) is the emitting process, where the black line represents the input phonon shape, and the red, blue and green lines represent the output phonon shapes when the coupling strength are at $g_{m}/2\pi =0.96MHz, 0.64 MHz$ and $0.32 MHz$ respectively with the same value of $\kappa =0.32MHz$. The overlap between the input pulse and the output pulse are $98.57\%$, $98.32\%$ and $94.38\%$ for the red, blue and green lines respectively.}\label{fig_robustness3}
\end{center}
\end{figure}

After absorbing and detecting the single phonon, we can apply a driving pulse $\Omega (
t) $ on the NVE, the phonon we have just absorbed will then be emitted.
The comparison of the input and output phonon pulse shape is shown in Fig.
3b with the $\kappa =0.32MHz$. The black line is the input phonon shape, the red, blue and green ones are the output phonon shape with $g_{m}/2\pi=0.96MHz$, $0.64MHz$ and $0.32MHz$
respectively. we can see that, the more stronger the coupling is, the larger
overlap it has between the input and output single phonon shape. When the
coupling strength $g_{m}/2\pi=0.96MHz$, the overlap can reach $98.57$. In the
first step, the NVE absorbs one phonon from the diamond or not. And then, we
can detecting wether the single phonon state was absorbed through the frequency shift of the phononic crystal. Finally, the NVE can emit the phonon with nearly
the same shape compared with the absorbed one. In this process, the single phonon
state has been measured without changing the phonon shape, which is a true
QND measurement. The process of emitting can act as a single phonon
source. The whole courses including absorbing, detecting and emitting a single phonon can also serve as a single phonon memory.

\section{Conclusion}

We have proposed a scheme to realize the QND single phonon state detecting and emitting based on the strain mediated interaction between the NVE and the
single phonon. By analyze the dynamical evolution of the real
system, we are able to calculate the fidelity of the absorbing process and
the overlap between the input and output phonon number state, both of which can reach a very
high value. The emitting process can act as a single phonon source and the whole courses including absorbing, detecting and emitting a single phonon can also serve as a single phonon memory. In future, the similar method may also be used to realize the QND measurement for arbitrary Fock states of phonons.

\begin{acknowledgments}
R.X.W. is supported by China Postdoctoral Science Foundation (2016M600999). Z.Q.Y. is supported by National Natural Science Foundation of China
(NSFC) (61435007 and 61771278), and Joint Fund of the Ministry of Education
of the People's Republic of China (MOE) (6141A02011604). G.L.L is supported by National Natural Science Foundation of China
(NSFC) (20141300566).
\end{acknowledgments}


\begin{thebibliography}{99}
\bibitem{DLCZ} L. M. Duan, M. D. Lukin, I. Cirac, P. Zoller, Nature \textbf{414}, 413 (2001).

\bibitem{Yin2007} Z. Q. Yin and F. L. Li, Phys. Rev. A 75, \textbf{012324} (2007).

\bibitem{Song2017} Chao Song, et al, arXiv:1703.10302.

\bibitem{QizheIn1} Q. Hou, W. Yang, C. Chen and Z. Yin, J. Opt. Soc. Am. B \textbf{33}, 2242 (2016).

\bibitem{WangIn4} X. Wang, A. Miranowicz, H. R. Li, F. Nori, Phys. Rev. A \textbf{93}, 063861 (2016).

\bibitem{VladimirIn6} , V. M. Stojanovic, M.Vanevic, E. Demler, and L. Tian, Phys. Rev. B \textbf{89}, 144508 (2014).

\bibitem{ConnellIn7} A. D. O'Connell, M. Hofheinz, M. Ansmann, et al., Nature, \textbf{464}, 7289 (2010).

\bibitem{SchuetzIn2} M. J. A. Schuetz, E. M. Kessler, G. Giedke, L. M. K. Vandersypen, M. D. Lukin, and J. I. Cirac, Phys. Rev. X \textbf{5}, 031031 {2015}.

\bibitem{Patricio2016} P. Arrangoiz-Arriola and A. H. Safavi-Naeini, Phys. Rev. A \textbf{94}, 063864 {2016}.

\bibitem{Martin2014} M. V. Gustafsson, T. Aref, A. F. Kockum, M. K. Ekstr$\ddot{o}$m, G. Johansson, and P. Delsing, science \textbf{346}, 6206 {2014}.


\bibitem{painter2015} S. M. Meenehan, J. D. Cohen, G. S. MacCabe, F. Marsili, M. D. Shaw, and O. Painter, Phys. Rev. X \textbf{5}, 041002 {2015}.



\bibitem{ManentiIn3} R. Manenti, A. F. Kockum, A. Patterson, T. Behrle, J. Rahamim, G. Tancredi, F. Nori, and P. J. Leek, arXiv:1703.04495.



\bibitem{Galland3} C. Galland, N. Sangouard, N. Piro, N. Gisin, and T. J. Kippenberg, Phys. Rev. Lett. \textbf{112}, 143602 (2014).

\bibitem{Matsuda5} O. Matsuda, O. B. Wright, D. H. Hurley, V. Gusev, K. Shimizu, Phys. Rev. B \textbf{77}, 224110 (2008).

\bibitem{Matsuda6} O. Matsuda, O. B.Wright, D. H. Hurley, V. E. Gusev, and K. Shimizu, Phys. Rev. Lett. \textbf{93}, 095501 (2004).

\bibitem{Yanay4} Y. Yanay and A. A. Clerk, New J. Phys. \textbf{19}, 033014 (2017).

\bibitem{Chai7} J. H. Chai, Y. Q. Lu, Physica B \textbf{291}, 292 (2000).

\bibitem{Bai8} X. Baia, T. A. Eckhausea, S. Chakrabartib, P. Bhattacharyab, R. Merlina, C. Kurdak, Physica E \textbf{34}, 592 (2006).

\bibitem{Woolley11} M. J. Woolley, A. C. Doherty, and G. J. Milburn, Phys. Rev. B \textbf{82}, 094511 (2010).

\bibitem{Ohm9} C. Ohm, C. Stampfer, J. Splettstoesser, and M. R. Wegewijs, Appl. Phys. Lett. \textbf{100}, 143103 (2012).

\bibitem{Neto10} O. P. deSaNeto, M. C. deOliveira, F. Nicacio, and G. J. Milburn, Phys. Rev. A \textbf{90}, 023843 (2014).

\bibitem{Haroche2007} S. Gleyzes, S. Kuhr, C. Guerlin, J. Bernu, S. Del$\acute{e}$glise, U. B. Hoff, M. Brune, J. M. Raimond and S. Haroche, nature \textbf{446}, 05589 (2007)


\bibitem{painter2009} M. Eichenfield, J. Chan, R. M. Camacho, K. J. Vahala, and O. Painter, Nature \textbf{462}, 78 (2009).

\bibitem{Ovartchaiyapong2014} P. Ovartchaiyapong, K. W. Lee, B. A. Myers, and A. C. B. Jayich, Nat. Comm. \textbf{5}, 4429 (2014).

\bibitem{Golter2016} D. A. Golter, T. Oo, M. Amezcua, K. A. Stewart, and H. Wang, Phys. Rev. Lett. \textbf{116}, 143602 (2016).

\bibitem{MacQuarrie2017} E. R. MacQuarrie, M. Otten, S. K. Gray, and G. D. Fuchs, Nat. Comm. \textbf{8}, 14358 (2017).

\bibitem{Rabl2008} P. Rabl, P. Cappellaro, M. V. G. Dutt, L. Jiang, J. R. Maze, and M. D. Lukin, Phys. Rev. \textbf{79}, 041302 (2008).

\bibitem{Rabl2010} P. Rabl, S. J. Kolkowitz, F. H. L. Koppens, J. G. E. Harris, P. Zoller, and M. D. Lukin, Nat. Phys. \textbf{6}, 602 (2010).

\bibitem{du2009} Z. Y. Xu, Y. M. Hu, W. L. Yang, M. Feng and J. F. Du, Phys. Rev. A \textbf{80}, 022335 (2009).

\bibitem{zhou2009} L. G. Zhou, L. F. Wei, M. Gao, and X. B. Wang, Phys. Rev. A \textbf{81}, 042323 (2010).

\bibitem{yin2013} Z. Q. Yin, T. C. Li, X. Zhang, and L. M. Duan, Phys. Rev. A \textbf{88}, 033614 (2013).

\bibitem{ma2016} Y. Ma, Z. Q. Yin, P. Huang, W. L. Yang, and J. F. Du, Phys. Rev. A \textbf{94}, 053836 (2016).

\bibitem{cai2017} K. Cai, R. X. Wang, Z. Q. Yin, G. L. Long, Sci. Chi. \textbf{60}, 070311 (2017).

\bibitem{Bennett2013} S. D. Bennett, N. Y. Yao, J. Otterbach, P. Zoller, P. Rabl, and M. D. Lukin, Phys. Rev. Lett. \textbf{110}, 156402 (2013).

\bibitem{wang2016x} D. A. Golter, T. Oo, M. Amezcua, I. Lekavicius, K. A. Stewart, and H. Wang, Phys. Rev. X \textbf{6}, 041060 (2016)

\bibitem{wang2016l} D. A. Golter, T. Oo, M. Amezcua, K. A. Stewart, and H. Wang, Phys. Rev. Lett. \textbf{116}, 143602 (2016)

\bibitem{maze2011} J.R. Maze, A. Gali, E. Togan, Y. Chu, A. Trifonov, E. Kaxiras, and M. D. Lukin,  New J. Phys. \textbf{13}, 025025 (2011).

\bibitem{doherty2011} M. W. Doherty, N. B. Manson, P. Delaney, and L. C. L. Hollenberg, New J. Phys. \textbf{13}, 025019 (2011).







\bibitem{Braginsky14r1} V. B. Braginsky, and F. Y. Khalili, Quantum Measurement (ed. Thorne, K. S.) Chs IV and XI (Cambridge Univ. Press, Cambridge, UK, 1992).

\bibitem{Grangier14r2} P. Grangier, J. A. Levenson, and J. P. Poizat, Nature \textbf{396}, 537 (1998).

\bibitem{Nogues14r3} G. Nogues, et al., Nature \textbf{400}, 239 (1999).

\bibitem{Manson1r49} N. B. Manson,  J. P. Harrison, and M. J. Sellars, Phys. Rev. B \textbf{74}, 104303 (2006).

\bibitem{Saito12} S. Saito, X. Zhu, R. Ams\"{u}ss, Y. Matsuzaki, K. Kakuyanagi, T. Shimo-Oka, N. Mizuochi, K. Nemoto, W. J. Munro, and K. Semba, Phys. Rev. Lett. \textbf{111}, 107008 (2013).

\bibitem{Rogers2008} L. J. Rogers , S. Armstrong , M. J. Sellars and N. B. Manson, New J. Phys., \textbf{10}, 103024 (2008)



\bibitem{Yin2015a} Zhang-qi Yin, Zhao Nan, Tongcang Li, Science China Physics, Mechanics \& Astronomy \textbf{58}, 050303 (2015),

\bibitem{Duan2003}   L. M. Duan, A. Kuzmich, and H. J. Kimble, Phys. Rev. A \textbf{67}, 032305 (2003).

\bibitem{Yin2015} Zhang-qi Yin, W. L. Yang, Luyan Sun, L. M. Duan Phys. Rev. A \textbf{91}, 012333 (2015).


\bibitem{Cirac13} J. I. Cirac, P. Zoller, H. J. Kimble, and H. Mabuchi, Phys. Rev. Lett. \textbf{78}, 3221 (1997).

\bibitem{James2} D. F. V. James and J., Canadian Journal of Physics,  \textbf{6}, 625 (2007).

\bibitem{Robledo} L. Robledo, H. Bernien, V. D. S. Toeno, R. Hanson, New J. Phys. \textbf{13}, 119 (2010).

\bibitem{Gill2013} N. Bar-Gill, L.M. Pham, A. Jarmola, D. Budker, and R.L. Walsworth, Nat. Commun. \textbf{4}, 1743 (2013).

\bibitem{Tao2014} Y. Tao, J. M. Boss, B. A. Moores, and C. L. Degen, Nat. Commun. \textbf{5}, 3638 (2014).






\end{thebibliography}
\end{document}